\documentclass[aps,reprint,amsmath,amssymb,graphicx,bibliography]{revtex4-2}

\usepackage{graphicx}% Include figure files
\usepackage{xcolor}
\usepackage{dcolumn}% Align table columns on decimal point
\usepackage{bm}% bold math
\usepackage{color}
\usepackage{enumitem} %for itemizing
\begin{document}
\color{black}

\title{Use of Superfluid Helium to Observe Directionality of Galactic Dark Matter }

\author{G.~M.~Seidel}
\email{george\_seidel@brown.edu}
\affiliation{Department of Physics, Brown University, Providence,
  Rhode Island 02912, USA}

\author{C.~Enss}
\email{christian.enss@kip.uni-heidelberg.de}
\affiliation{Kirchhoff Institute for Physics, Heidelberg University, 69120 Heidelberg, Germany}
%8 3 Institute for Data Processing and Electronics (IPE), Karlsruher Institut für Technologie}

\date{\today{}}

\begin{abstract}

The quasiparticle propagation away from the track of a highly ionizing particle in superfluid helium at low temperatures has previously been shown to exhibit anisotropy. We discuss the mechanism responsible for this behavior and show that it occurs for nuclear scattering by dark matter for recoil energies down to a few keV, and perhaps lower. This makes it possible to extend  WIMP searches with interaction cross sections that reach into the neutrino floor in a meaningful energy range.

%Recent progress in the use of superfluid helium at low temperatures as a potential detector for dark matter, together with developments in supporting technologies, has raised the possibility of a number of interesting variations in detector design as well as suggestions for new applications.  The superfluid can be an excellent directional  dark matter detector for recoil energies about a few keV. A means of enhancing the efficiency of detecting recoil energies down to 1 meV is also discussed.

\end{abstract}

%\pacs{81.05.Uw,68.37.-d,73.20-r}

\maketitle
\section{Introduction}
Interest in superfluid helium as a medium in which to search for rare neutrino events began 35 years ago\cite{Lanou_1987}  because of two beneficial properties of the superfluid. One, its purity. Nothing is soluble in liquid $^4$He except for the rarer isotope $^3$He. And two, it is possible to extract from the bulk liquid, at very low temperature, the thermal energy deposited by a scattering event using the process of quantum evaporation\cite{Balibar_2016, Wyatt_1992}. There have subsequently been many papers discussing such topics as its use for detection of dark matter\cite{Lanou_1996,Guo_2013}, experimental studies of energy deposition in liquid helium\cite{Porter_1994,Bandler_1995a,Bandler_1995b,Adams_1998,McKinsey_1999,Adams_2001,McKinsey_2004,Carter_2017}, different possible designs of detectors\cite{Huang_2008,Hertel_2019}, calculations of expected discrimination between nuclear and electron recoils\cite{Ito_2013,Seidel_2014}. Recently, important progress has been reported\cite{McKinsey_2022} of results on a helium detector, with much improved calorimeter/sensors, such that it is now technically capable of contributing to the search for low mass WIMPs. Two collaborations have been formed with the aim to use superfluid helium for dark matter searches \cite{Hertel_2019,Krosigk_2022}.\\

As  growing regions in cross-section/mass space for WIMPs are  being excluded  by the many ongoing dark matter searches, there have been numerous proposals, see Ref.\cite{Baracchini_2022}, to extend the search region into the neutrino fog by using the directionality of nuclear recoils from galactic WIMPs for discrimination.  Perhaps the most developed of these proposals is CYGNUS \cite{Baracchini_2022,Vahsen_2021}, which uses helium gas at 1 bar and room temperature as the target material in a time projection chamber (TPC) for detection. In order to have sufficient target mass 100 TPCs each having a volume of 10 m$^3$ are proposed. The same helium mass can be contained as a liquid in a  1 m$^3$ volume.  And in the liquid a nuclear scattering event having an energy greater than a few keV is expected to have a  roton emission that is spatially  highly anisotropic thereby providing a direct means of determining the track direction.\\

The angular distribution of rotons propagating away from the tracks of alpha particles in liquid helium below 0.1 K with a mean energy of 3.3 MeV have been shown to be highly asymmetric\cite{Bandler_1995b}. The quasiparticle flux perpendicular to the track is 4 times larger than that  propagating parallel to track direction. This anisotropy results from the fact that the initial density of quasiparticles, principally rotons from phase space arguments, along the track is very high. The rotons interact and form a hot thermalized cloud. A roton escapes the columnar cloud  into the cold, roton-free surrounding liquid after its last scatter with a velocity  that is preferentially oriented perpendicular to the columnar track. We discuss below in Section II that  this anisotropy can, with confidence,  be expected to hold for tracks having recoil energies down to 10 keV. This confidence is based on  our understanding of the energy deposition in superfluid helium by nuclear recoils. For recoil energies below several keV where electronic stopping power dominates over nuclear stopping power in helium, the situation is changed. While directionality is expected to occur at these low energies this remains to be demonstrated. This is discussed separately in Section III.   \\

The use of quantum evaporation to detect quasiparticles in helium has limitations in determining asymmetry. The observation of track anisotropy\cite{Bandler_1995b} used a collimated source, the orientation of which could be varied with respect to the liquid surface. Since, for kinematic requirements, a roton must be incident on the free surface within a critical cone of 20$^\circ$ degrees (Snell's law) to evaporate an atom, very little information about directionality is transferred to an array of calorimeters detecting the evaporated helium atoms.  For superfluid helium to become a useful directional detector of WIMPS, rotons should be measured in the liquid, which requires heat transport from liquid helium into a solid.  We  discuss this topic in Section IV. \\

\section{Directional Detection in superfluid helium above 10 keV recoil energy}

For liquid helium to be useful as a directional detector for dark  matter, roton anisotropy must persist for nuclear scattering with recoil energies far less than 3 MeV\cite{Bandler_1995b} where it has been observed. To make an evaluation of the likelihood that this will occur,  requires a  discussion of the mechanisms by which quasiparticles are produced in high and in low energy recoils. \\

When a WIMP, in the process of interacting with a helium atom, transfers more than 10 keV, that energy is predominantly kinetic energy of the scattering atom. This energy is rapidly transformed into that of  ionizations and excitations of other atoms along the path of the primary. Since collisions between helium atoms has been extensively studied in atomic beam experiments, it is possible to analyse in detail the fraction of energy that results in ionization and excitations and which subsequently results in rotons, UV and longer wavelength radiation, and long-lived  (13 s) triplet excimers\cite{Ito_2013}. An important check on the reliability of these calculations is provided by the excellent agreement that is obtained for the stopping power of alpha particles in helium between 10 keV and 5 MeV calculated from measured atomic scattering processes and the accepted tabulated values\cite{Nist}.\\ 

At high energies ($>$10 keV) the primary scattered helium atom loses its kinetic energy by exciting electrons on atoms along its trajectory. The stopping power is predominately electronic. Momentum transfer is small and the track is linear. Charge transfer reactions do not affect this conclusion. Nuclear scattering makes virtually no contribution to quasiparticle creation. The average energy to produce an ionization event, W value = 43 eV, is a result of ionization (24.6 eV) plus electron recoil energy (8 eV) and electronic excitations (11 eV) which occur about half as frequently as ionizations\cite{Tenner_1963,Ito_2013}. Quasiparticles are produced by a number of different processes: stopping of the secondary electrons, excimer formation, non radiative quenching by Penning ionization, damping of rotational states of excimers and displacement of the two ground state helium atoms after having undergone a radiative transition to the unbound ground state. Altogether it is estimated that slightly more than 20  eV of the original 43 eV ends up as quasiparticles\cite{Ito_2013}, the principal uncertainty arising from parameters used  in calculating the Penning quenching reactions.\\

The quasiparticles are initially dispersed within a volume about the track determined by the distribution of the recoil electrons. The positive ion is initially located directly on the track and forms a solid-like "snowball" with an effective mass of $\sim$40 m$_{He}$ by the polarization of surrounding atoms. The electron bubble formed after the recoil electron has slowed sufficiently  has an effective hydrodynamic mass of 240 m$_{He}$. Hence, recombination results primarily from the motion of the positive ion, and quasiparticle generation is closely related to the electron distribution.\\

The electron distribution  has  been determined for secondary electrons about the tracks of energetic electrons emitted in superfluid helium by a $^{63}$Ni $\beta$ source \cite{Seidel_2014}.  This determination is made possible by measuring the recombination of geminate electron/ion pairs, in the absence of appreciable diffusion, as a function of an applied electric field.  The electrons were found to be defined spatially  by a Gaussian distribution about the track with a half width of $4\times 10^{-6}$ cm  and accompanied by a long, fat tail containing about 10\% of the total charge, with a density falling off as a high power of distance ($\sim r^{-8}$). Since electrons generated by a recoil helium atom are comparable in  energy to  produced by  betas from the Ni source, the charge distribution for a nuclear recoils  can be taken as that found for the beta source.\\

The single differential cross section for scattering of electrons by helium  is peaked at very low energies, falling off 2 orders of magnitude between 10 and 100 eV\cite{Manson_1975}. For both types of incident projectiles, electrons and nuclear recoils, the secondary electrons are expected to have mean energies the order of 10 eV. The electron distribution and hence the quasiparticle distribution about any track, for which the  energy loss is dominated by electron ionization, is primarily contained within a radius of $ 4\times 10^{-6}$ cm.\\

Whether the track of a nuclear recoil will exhibit anisotropic roton propagation  depends upon the ratio of the track length to its diameter and upon the initial density of quasiparticles. At a recoil energy of 10 keV the track length is $\sim 1.4\times 10^{-4}$ cm\cite{Ito_2013}, about 20 times the  diameter of the electron distribution.   The quasiparticles initially form a cylindrical cloud with a density greater than   $10^{20}$ cm$^{-3}$. The cross section for roton-roton scattering has been found to vary between 50 and 150 $\times 10^{-14}$ cm$^2$ \cite{Forbes_1990, Bauer_1992} so that quasiparticles  scatter and thermalize prior to escaping the cloud. Upon escaping they will do so preferentially with velocities perpendicular to the axis of the track. The asymmetry  may vary somewhat depending on recoil energy and track length. Simulations have yet to be performed to assess this difference. \\

The energy deposition along the track from an electron, while considerably less dense than for a nuclear recoil,  can still reach a quasiparticle concentration sufficient, in principle, for anisotropic roton propagation. No such anisotropy was observed using collimated  monoenergetic 364 keV electrons from the capture reaction in $^{113}$Sn \cite{Adams_2001}. The isotropy of the signal is presumably the result of large angle scattering of the electron projectile and its highly non linear track.\\

 \section{Directionality below 10 keV}
 For a 1 keV nuclear recoil the situation is somewhat different than for 10 keV and higher energies. The stopping power for a helium atom shifts from nuclear to electronic below 7 eV. And at this energy and below  the cross section for ionization is more than an order of magnitude less than that for excitation, primarily to the $2^1$P state\cite{Ito_2013}. On average only a few percent of the initial recoil energy ends up in an ionizing event. The same is true for the creation of long-lived triplet excimers as their creation is principally the result of ionization and recombination. \\
 
  In contrast to rotons resulting from ionizations,  quasiparticles produced as a result of electronic excitations are tightly confined  within a radius no more than a few times $10^{-7}$ cm about the track of the initiating atom.  The excited atom does not stray from the track.  Also, any quasiparticles produced by atomic collisions are highly localized about the track as well \cite{You_2022}. The track length, based upon Lindhard\cite{Lindhard_1963}, is the order of $10^{-5}$ cm. The track would have a high length to diameter ratio, more than 10.  While the total energy converted into  rotons is lower, the much smaller cylindrical column in which they are contained results in  their having a extremely high density  and thermalizing before propagating away anisotropically. \\
 
Whether the asmmetry of such a small energy deposition can be measured in superfluid helium is an open question discussed below.\\

\section{Heat Transport through liquid helium/solid surfaces}
The transport of thermal energy through interfaces, in particular the liquid helium/solid surface, has been of interest for many years. Ref. \cite{Swartz_1989}  furnishes a review of early work on the subject with Ref. \cite{Chen_2022} providing a very recent summary. Unfortunately, there is little work, experiment or theory, that is related  to energy transport through a surface by incident rotons. The only theoretical discussion of rotons interacting with surfaces is that of the Kharkov group\cite{Adamenko_2008,Tanatarov_2010}  and is based on the assumption of  an idealized smooth, flat surface. This assumption is known to work well for long wavelength phonons, in keeping with the Khalatnikov acoustic mismatch model and thermal conductance measurements at very low temperatures, $T<0.1$ K, but has been amply demonstrated experimentally not to hold  at higher temperatures\cite{Olson_1994,Ramiere_2016}, the  conductance  being orders of magnitude higher than predicted theoretically.
Olson and Pohl\cite{Olson_1994}  found the conductance through a polished surface agreed with acoustic mismatch theory at low temperature (0.1\,K) but was 20 times higher than predicted at 2 K for the polished surface and more that 100 times higher for a roughened surface. There is no agreed upon explanation of this behavior. A number of possible causes have been proposed that can explain some experimental results but not others. These include surface roughness, imperfections or oxide layers resulting in inelastic scattering, surface modes opening new channels for energy transfer, and enhanced phonon transmission because of the boundary layer of solid helium on the surface. None of these studies, mainly performed above 1 K where rotons are the principal thermal carrier in helium, discuss the possible role of rotons in interfacial energy transfer.\\

Knowledge of what  occurs when a roton interacts with a solid surface is important in designing and evaluating the performance  of a superfluid helium dark matter detector. The only experimental information currently available appears to be an estimate by Bandler\cite{Bandler_1995a} of the reflection of rotons travelling ballistically at low temperature. Using rotons produced by alpha particles stopped in the liquid and the detection of atoms subsequently evaporated from the surface, he demonstrated that the reflection from a variety of surfaces is diffusive with an estimated probability of 30\%. \\

\section{Directional Detection in superfluid Helium}

It is not our purpose here to detail what a  directional detector of dark matter based on superfluid helium would look like.  Rather the object of this discussion is to point out that current technology is capable of instrumenting such a directional detector. However,  as noted above,  the current means of detecting quasiparticles produced in the liquid by using quantum evaporation, is ill suited for discriminating directional tracks.  Something the order of $4\pi$ sensor coverage in the liquid would be desirable.\\

Low temperature calorimetric sensors, read out by SQUIDs, presently have the energy sensitivity required to measure a 10 keV recoil in  1 m$^3$ of liquid helium. Calorimetric wafers having an area of 50 cm$^2$ based on transition edge sensors (TES)\cite{Fink_2021} and by  metallic magnetic calorimeters (MMC)\cite{Gray_2016,Poda_2021}  currently have energy thresholds of 3 eV. As an example, a 1 m$^3$ sphere having a surface area of 5 m$^2$, would require 1000 such 50 cm$^2$ calorimeters for complete coverage.  For a 10 keV recoil the average energy absorbed by a single wafer would be $ 10 \rm{eV}\times .5\times.7= 3.5$ eV. The 10 eV is the  energy of 10 keV distributed uniformly over the 1000 calorimeters. The factors of .5 and .7 account, respectively; for the fraction of the recoil energy in the form of quasiparticles and for reflection at the surface.\\ 

A superfluid detector with such an array of energy sensors would certainly be able the detect anisotropic roton signals from energy recoils above 10 keV. There is, however, one caveat to this discussion of signal detection. These large-area wafer/calorimeters have not been shown to operate within superfluid helium, for the reason that there has not been the need for them to do so. A concern is that much of the thermal energy deposited in the wafer could be transferred back into the liquid and will not reach the temperature-sensitive element of the sensor. Should this be a problem with present sensor designs, one solution could be the development of surfaces that quickly convert the adsorbed energy into modes that have a low probability of transmission back into the liquid, e.g.,  into long wavelength phonons by down conversion or into trapped superconducting quasiparticles. While this issue would appear to have a solution, it needs to be addressed before serious consideration is given to the design  of a superfluid helium-based  directional dark matter detector. \\

 The readout of large arrays of SQUID-based sensors by multiplexing, while not pro forma, is a well established technology, the intricacies of which depend on a large number of parameters associated with the particular application. An array of a thousand pixels is a quite modest size. The special features of the sensors and how they might be fabricated is not germaine to this discussion of the properties of superfluid helium that make it a medium for anisotropic track detection. Some of their properties and how they might be used, however, do warrant comment.\\ 

The sensors do not have to be fast, as SQUID sensors go. Background event rates in WIMP detectors of comparable size are the  order of one per second in well-shielded underground environments. A response time of a few tens of microseconds would be adequate for timing and event location, as the roton velocity is the order of 20 m/s. It would also allow for the discrimination of rotons arriving later, not having been absorbed on hitting a surface but reflected.\\

The hypothetical geometry of a helium detector, discussed above, a 1000 m$^3$ volume  the surface of which is covered by sensors, is only the simplest example of how such a detector might be configured. Presumably the mass of helium should not be much less as the number of WIMP scatters would only be a few per year in 1000 m$^3$ of liquid for scattering  cross sections in the neutrino fog. However, one could consider placing arrays of sensors within the liquid, dividing the total volume into smaller compartments. Such an option, depending on design, opens up a number of possible improvements in function and performance,  many of which might include trade offs such a the  readout of larger arrays of sensors. The helium volume and sensor placement could be designed to decrease the  threshold recoil energy sensitivity to 1 keV or lower, or rather to enhance the discrimination anisotropic tracks at higher energy recoils, or alternatively to change the design parameters of  the calorimetric sensors. None of these possibilities, and others, are not within the scope of this paper.\\

In summary, superfluid helium is an excellent target material for the directional detection of galactic dark matter searches should the need arise to push parameter space into the neutrino fog.\\

\section*{acknowledgments}
We have benefited from  informative conversations with Scott Hertel.

%\clearpage

%\bibliography{comments}

\end{document}